\documentclass[10pt,journal,compsoc]{IEEEtran}

\ifCLASSOPTIONcompsoc
  \usepackage[nocompress]{cite}
\else
  \usepackage{cite}
\fi

\ifCLASSINFOpdf

\else
\fi

\usepackage[T1]{fontenc}
\usepackage{txfonts}
\usepackage[pdftex]{hyperref}
\usepackage{color}
\usepackage{siunitx}
\usepackage{booktabs}
\usepackage{textcomp}
\usepackage{multirow}
\usepackage[caption=false]{subfig}
\usepackage{microtype} 
\usepackage{ccicons}  
\usepackage{mathptmx}
\usepackage{graphicx}
\usepackage{times}
\usepackage{enumitem}
\usepackage{tabularx}
\usepackage[nopar]{lipsum}
\usepackage{sparklines}

\usepackage{cite}

\begin{document}

\title{Evaluation of Visualization by Demonstration and Manual View Specification}

\author{Bahador~Saket and
        Alex~Endert
\IEEEcompsocitemizethanks{\IEEEcompsocthanksitem Bahador Saket and Alex Endert are
with Georgia Tech.\protect\\
E-mail: \{saket, endert\}@gatech.edu.

}
\thanks{Manuscript received April 19, 2005; revised August 26, 2015.}}

\markboth{Journal of \LaTeX\ Class Files,~Vol.~14, No.~8, August~2015}%
{Saket \MakeLowercase{\textit{et al.}}: Evaluating Interactive Graphical Encodings for Data Visualization}

\IEEEtitleabstractindextext{%
\begin{abstract}
We present an exploratory study comparing the visualization construction and data exploration processes of people using two visualization tools, each implementing a different interaction paradigm. One of the visualization tools implements the manual view specification paradigm (\href{https://github.com/vega/polestar}{Polestar}) and another implements the visualization by demonstration paradigm (\href{https://github.com/BahadorSaket/VbD}{VisExemplar}). Findings of our study indicate that the interaction paradigms implemented in these tools influence: 1) approaches used for constructing visualizations, 2) how users form goals, 3) how many visualization alternatives are considered and created, and 4) the feeling of control during the visualization construction process.
\end{abstract}

}

\maketitle

\IEEEdisplaynontitleabstractindextext

%
\IEEEpeerreviewmaketitle

\ifCLASSOPTIONcompsoc
\IEEEraisesectionheading{\section{Introduction}\label{sec:introduction}}
\else
\section{Introduction}
\label{sec:introduction}
\fi

\IEEEPARstart{V}{isual} representation and interaction are two main components of visualization tools~\cite{yi2007toward, sedig2004need}. The main focus of visual representation is mapping data values to graphical representations and rendering them on a display in an effective manner. Interaction provides users the ability to change the parameters of the system to construct and change visual representations and interpret the resulting views~\cite{yi2007toward, dix1988starting}. While the goal of most visualization tools is to accommodate visualization construction and data exploration, they may use different \textit{interaction paradigms} (defined in Table~\ref{table:terminology}) to accomplish this.

A commonly used interaction paradigm in most visualization tools is \textit{manual view specification}. Tools implementing manual view specification often require users to manually specify the desired mappings
through GUI operations on collections of visual properties and data attributes that are presented visually on control panels. For instance, to create a scatterplot, users must specify the visualization technique, then select data attributes to map onto the axes, and finally map any additional encodings used (e.g., size, color, etc.) to the desired attributes. In this paradigm, users are responsible for specification of the mappings, while the system computes the resulting view. Manual view specification frequently used in successful visualization tools such as Spotfire~\cite{SpotFire}, Tableau~\cite{Tableau}, and many more.

Saket et al. recently proposed an alternative interaction paradigm for visualization construction and data exploration called \textit{visualization by demonstration}~\cite{saketVbD}. This paradigm advocates for a different process of visualization construction. Instead of specifying mappings between data attributes and visual representations directly, visualization by demonstration lets users demonstrate partial mappings or changes to the output (the visualization). From these given demonstrations, the system interprets user intentions, and recommends potential mappings. Visualization by Demonstration is inspired by previous work that indicated the effectiveness of letting people create spatial representations of data points manually, without the need to formalize the mappings between the data and the spatial constructs created~\cite{huron2014constructive, andrews2010space,shipman1999formality}.

\begin{figure}
\centering

  \includegraphics[width=\columnwidth]{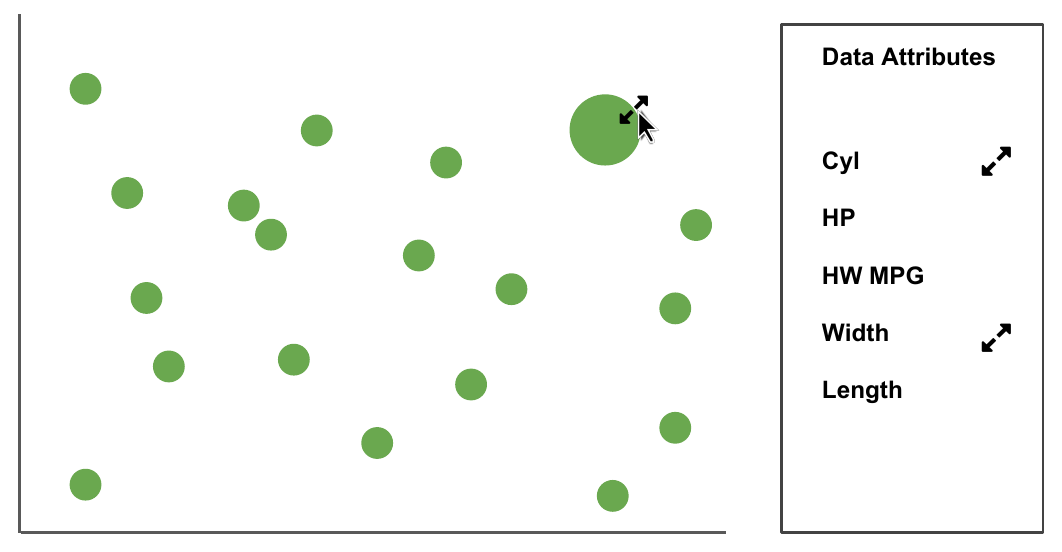}
\caption{A user directly interacts with a point by making its size larger to demonstrate the interest in mapping a data attribute to size. Based on the given demonstrations, the system extracts an appropriate set of data attributes that can be mapped to size and suggest them.}~\label{fig:VBD}
\end{figure}

For example, in a scatterplot, instead of mapping a data attribute to color through control panels (manual view specification), in visualization by demonstration users could color one or more data points to convey their interest in mapping color to a data attribute. Alternatively, users may be less familiar with the attributes of the dataset, and thus the demonstrations help them discover attributes that create these demonstrated mappings. In response to this demonstration, the system recommends a set of appropriate data attributes that can be mapped to color. From there, users could continue to construct their visualization by resizing data points to convey their interest in mapping size to a data attribute. In response, the system would extract data attributes that can be mapped to size and suggest them (see Figure~\ref{fig:VBD}). This form of analytic discourse would continue in a similar manner. This highlights the overall difference between these two interaction paradigms -- providing demonstrations on data items in visualizations to create mappings vs. specifying the mappings in external controls.

\begin{table*}
\footnotesize
  \caption{Definitions of terms used in this paper. } \label{table:terminology} \vspace{-0.5em}
  \centering
  \begin{tabular}{p{0.2\textwidth}p{0.75\textwidth}}
    \toprule
      {\sc Term} & {\sc Definition}\\
    \midrule
     Interaction Paradigm & We define the term interaction paradigm in information visualization to refer to the process of how visualization construction and exploration is fostered in a visualization tool. \\ \\
     Mapping & Assignment of a data attribute to a visual property such as axes. \\ \\
     
     Visual Demonstration & User-generated visual output showing a partial visual result, typically performed by manipulating the graphical encodings used in the visual representation. For example, coloring a data point to indicate interest in mapping a data attribute to color of all data points.\\ \\
     
     Visualization by Demonstration & An interaction 
     paradigm for constructing visualizations that allows users to provide visual demonstrations of incremental changes or partial mappings. The systems implementing visualization by demonstration then recommend potential transformations or mappings from the given demonstrations.\\ \\
     
     Manual View Specification & An interaction paradigm for constructing visualizations that requires users to first specify the desired mappings between data attributes and visual properties through GUI operations (often presented visually via control panels). The system then generates the resulting views based on the specified mappings. \\

    \bottomrule
 \end{tabular}
\end{table*}

Despite the fact that both interaction paradigms enable users to construct visualizations through an iterative process, there are fundamental differences between the design considerations that go into each of these paradigms. \textbf{First}, the \textit{``visualization process model''}~\cite{Chi:interactionmodel} used in each of the interaction paradigms is different. \textbf{Second}, each interaction paradigm requires users' involvement in a different way. \textbf{Third}, each interaction paradigm uses a different level of abstraction. We elaborate on these differences in Section~\ref{sec:motivation}.

Many visualization tools have implemented the manual view specification paradigm. These tools have been successful in easing the processes of visualization construction and data exploration~\cite{grammel2010information}. They allow users to interactively change parameters to construct visualizations and explore data instead of using programming. However, when new interaction paradigms such as visualization by demonstration are created, it raises a number of intriguing questions including: \textit{How do interaction paradigms enable different visualization construction processes?} \textit{How effective are each of these interaction paradigms for specific tasks?} \textit{Which interaction paradigms do people prefer when constructing visualizations and exploring data?} Understanding the differences and trade-offs between various interaction paradigms and how they are used for specific tasks can help designers and developers make informed decisions about adapting these paradigms in visualization tools.

In this paper, we take a step towards gaining a better understanding of the trade-offs between these two interaction paradigms. In particular, we examine trade-offs between two tools, one implementing manual view specification (Polestar) and another implementing visualization by demonstration (VisExemplar). We conducted an exploratory study to investigate how these tools affect strategies that people follow while exploring their data, the common patterns that appear, design choices, and the challenges people encountered using each tool.

The main contributions of this paper are: 1) a characterization of visualization processes by comparing two tools with different interaction paradigms; 2) a discussion of how the underlying interaction paradigm used in each tool influences the visualization construction process; and 3) results of our study which show trade-offs between the two studied interaction paradigms. 

\section{Motivation}
\label{sec:motivation}
\begin{figure}
\centering
      \includegraphics[width=0.96\linewidth]{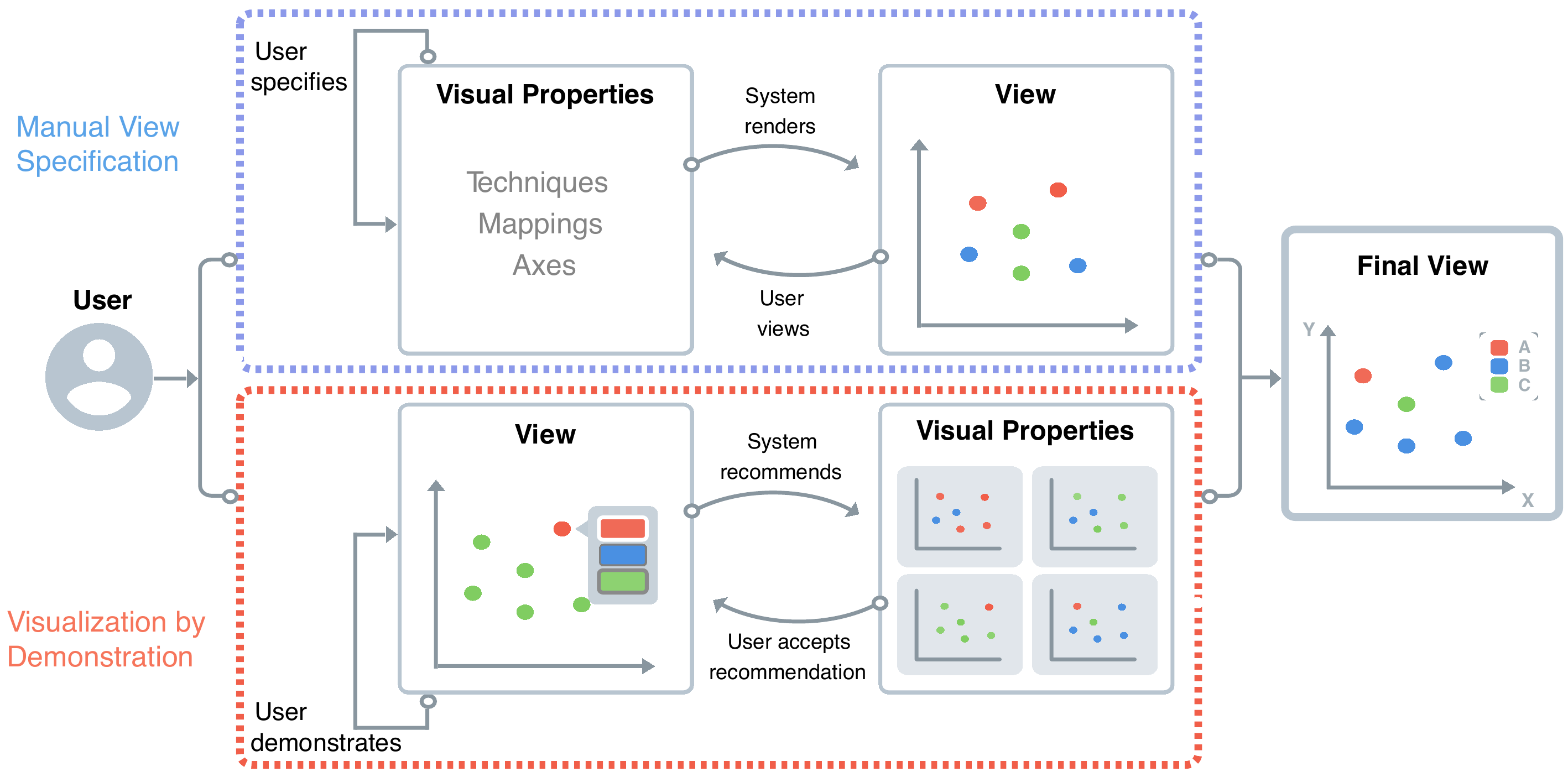}
  \caption{Visualization process using the manual view specification and visualization by demonstration paradigms.}~\label{fig:InteractionModel} \vspace{-0.8em}
\end{figure}

Manual view specification and visualization by demonstrations are two different interaction paradigms. 
Thus, their design differences (shown in Figure~\ref{fig:InteractionModel}) have direct implications on the visualization construction process of users. 
Below, we discuss these differences and how they motivate the importance of studying and understanding their impact. \\



\noindent\textbf{Each interaction paradigm uses a different level of abstraction.} In visualization by demonstration, users start at a low level of abstraction by manipulating the visual glyphs representing the data points to provide visual demonstrations of incremental changes. This is similar to the idea of constructive visualization~\cite{huron2014constructive}, which is defined as \textit{``the act of constructing a visualization by assembling blocks, that have previously been assigned a data unit through a mapping.''} In contrast, in manual view specification users start at an attribute level by specifying the overall mapping between attributes and visual encodings first. Tools implementing the manual view specification usually are not designed to allow the manipulation of the visual representation at an individual data point level. \\

\noindent\textbf{Each interaction paradigm uses a different \textit{``visualization process model"}~\cite{Chi:interactionmodel}.} The visualization process model in this case is defined as the process (or pipeline) of mapping data to visual properties~\cite{Chi:interactionmodel}. Manual view specification needs users to formally map data attributes to visualization properties prior to rendering the view ({\tt visual properties specification} $\rightarrow$ {\tt view rendering}). In contrast, visualization by demonstration first shows an initial visualization to users and allows them to provide visual demonstrations on the visualization. It then recommends potential mappings based on the given demonstrations ({\tt view rendering} $\rightarrow$ {\tt visual properties specification}  ). This difference is not only apparent for generating the initial view, but the interactive exploration process at subsequent steps also follows the same steps and sequences. 

\noindent\textbf{Each interaction paradigm enables different methods for user feedback and involvement.} In manual view specification, users are responsible for specifying data mappings and the system is responsible for rendering the view based on the specified mappings. In contrast, in visualization by demonstration, users are responsible for visually demonstrating their intentions and the system is responsible for recommending appropriate mappings based on the given demonstration (as well as rendering a view once a recommended mapping is chosen).


While interactions in visualization tools embodying a specific interaction paradigm can be implemented in various ways, the underlying interaction paradigm remains the same. For example, some of the tools that adopt manual view specification need users to assign data attributes to visual properties through dragging and dropping attributes to shelves (e.g., similar to Tableau), while others allow the specification through opening a drop-down menu to assign data attributes to visual properties. In both cases, the underlying interaction paradigm (manual view specification) requires users to manually assign data attributes to visual properties prior to rendering the view.

Similarly, interactions in tools embodying visualization by demonstration can be implemented in different ways. For example, tools might enable users to resize a data point by dragging a small handle on the perimeter of a glyph representing the data point or by dragging a slider that can be popped up upon right clicking the glyph. Regardless of how these interactions are implemented to enable users to provide demonstrations, the underlying interaction paradigm (visualization by demonstration) requires the system to interpret users' intentions and recommend potential mappings based on the given demonstrations.

\section{Related Work}

There are several visualization process models that explain different steps users follow to construct visualizations~\cite{card1999readings, Chi:interactionmodel, Carpendale:thesis, ware2010visual}. One of the well-known models is Card et al's ``reference model''~\cite{card1999readings}. 
The reference model explains four steps that users often follow to construct visualizations: \textbf{Raw Data Transformation}, \textbf{Data Table Transformation}, \textbf{Visual Properties Specification}, and \textbf{View Rendering}. Chi and Riedl~\cite{Chi:interactionmodel} proposed Data State Model of information visualization processes that extends the reference model proposed by Card et al. by allowing for multiple pipelines. Another variation of Card et al.'s reference model was introduced by Carpendale~\cite{Carpendale:thesis} that adds \textbf{Presentation Transformation} step before the view rendering. Despite the minor differences between these models, most of these models place \textit{visual properties specification} before \textit{view rendering}.
This implies that interaction paradigms that realize these models often ask users to assign data attributes to visual properties, then have systems render the specified views.

\subsection{Manual View Specification}

The manual view specification paradigm also applies the same visualization process model by requiring users to map data attributes to visual properties prior to rendering the view. Existing interactive visualization tools implementing this paradigm such as MS Excel, SpotFire~\cite{SpotFire}, Tableau~\cite{Tableau}, and Polaris~\cite{polaris:infovis00} require users to specify data mappings before rendering the view. This is how the visualization construction process is enabled at every step. For instance, to create a scatterplot, users must specify the point visualization technique, then select data attributes to map onto the x and y axes. The system then generates a scatterplot with all the specified characteristics.

Another set of tools that implement manual view specification is visualization authoring tools such as Lyra~\cite{2014-lyra}, iVisDesigner~\cite{ren2014ivisdesigner}, iVoLVER~\cite{Mendez:2016}, and Data-Driven Guides~\cite{kim2017data} that enable construction of flexible custom visualizations. These tools also require users to specify visualization properties prior to rendering the final view. However, these tools take a different approach from tools such as Tableau or MS Excel. Visualization authoring tools enable users to draw or import graphical elements. They then require users to manually establish mappings between data and the graphical elements that represent them prior to rendering the final view.

\subsection{Visualization by Demonstration}
A ``by-demonstration'' approach has been applied to a wide range of applications in computing before coming to data visualization. In computer programming, programming by demonstration~\cite{cypher1993watch} enables users to generate code by providing visual demonstrations of some intended result. The user and the system continue to collaborate, using further demonstrations or direct edits to incrementally improve the final code. Other domains that have successfully used the ``by demonstration'' approach include 3D drawing by demonstration~\cite{Igarashi:drawing}, data cleaning by demonstration~\cite{Lin:dataCleaning}, interactive database querying by demonstration~\cite{Zloof1975QueryBE} and more. 

A by-demonstration approach has also been used in visual analytic systems. For example, Kandel et al. showed how data wrangling and cleaning in spreadsheets can be done by demonstrating rules and filters~\cite{Kandel_wrangler}.
Dimension reduction and clustering models can also be guided by demonstrating group membership and relative similarity between data points~\cite{kwon2017axisketcher, endert2012semantic, endert2011observation, kiminteraxis}.



We previously presented visualization by demonstration~\cite{saketVbD}. In this paper, we presented a system implementing visualization by demonstration that shows an initial visualization to users and allows them to provide visual demonstrations of partial mappings or changes to the output (the visualization). From the given demonstrations, the system recommends a set of transformations. For each demonstration, the system first checks how the current state of the visualization and mappings should be transformed to create meaningful visual representations that match the demonstration. The system then recommends a set of possible transformations. Transformations can be a set of potential data attributes that can be mapped to different visual properties (e.g., assigning a data attribute to size of the data points), a set of visualization techniques that can be switched (e.g., transforming from a scatterplot to a bar chart), and other changes to view without changing underlying mappings (e.g., sorting a bar chart). The user and
the system continue to collaborate, incrementally producing more demonstrations and refining the transformations, until the most effective possible visualization is created.
Therefore, there is a growing interest within the visual analytics community to develop systems that use by-demonstration, which further motivates the importance to study and compare to existing interaction paradigms.




\subsection{Sketch-based Visualization Construction}

A related line of existing work within the visualization community has examined how sketching or drawing can be used as an approach to allow users to specify their intentions~\cite{lee2013sketchstory, kwon2017axisketcher, sketchbydemo, walny2012understanding, chao2010poster}. For instance, SketchStory~\cite{lee2013sketchstory} is a system that allows users sketch out an example icon and visualization axis. The system then interprets user interactions and completes the visualization with new visual properties. 
The sketching-based paradigm relies on digital ink and ink recognition. In addition, it relies on users exemplifying their intentions via drawing. However, there are commonalities with visualization by demonstration in the sense that both allow users to specify their desired goal instead of manually parameterizing the visual properties.

\subsection{Previous Studies of the Interactive Visualization Process}
Previous studies exist that have investigated the visualization construction process using different tools (e.g., ~\cite{TangibleTiles, Mendez:2017,grammel2010information, SamuelThesis }). For example, Wu et al.~\cite{TangibleTiles} compared the bar chart construction process between MS Excel and a set of physical tiles. Their findings showed that the distribution of time spent and sequence of actions taken to construct visualizations are different depending on the tool. Unlike the study by Wu et al.~\cite{TangibleTiles}, we investigate trade-offs of visualization construction and data exploration in digital visualization tools. In another study, Mendez et al.~\cite{Mendez:2017} compared how visualization novices construct visualizations and make design choices using either top-down or bottom-up visualization tools. Their findings indicated trade-offs between the two tools and considerations for designing better interactive visualizations. While we also explore the interactive visualization construction and data exploration process, we particularly focus on investigating the trade-offs between two different interaction paradigms. In particular, we are interested in investigating how each of the two interaction paradigms (manual view specification and visualization by demonstration) influence the visualization process.

\section{Study Design and Choice of Visualization Tools}

To investigate trade-offs between the two interaction paradigms, we selected two tools that each embody one of the two interaction paradigms examined in this paper. We did not include visualization tools implementing interaction paradigms designed for touch-based or sketch-based interfaces because we found it difficult to control for external factors (different input devices and interfaces) that might affect the study. Additional studies are required to investigate interaction paradigms that use input devices other than mouse and keyboard, and interfaces other than desktop devices.


The main experiment used two visualization tools, VisExemplar~\cite{saketVbD} and Polestar~\cite{PoleStar}, which satisfied two requirements. One is that each had to clearly embody one of the two interaction paradigms examined in this paper. The other is that to have a fair comparison between these two paradigms, users had to be able to learn and use the system within the duration of the experiment. As each tool adopted only one of the two interaction paradigms, we compared VisExemplar, which incorporates visualization by demonstration and Polestar, which incorporates manual view specification. Although a variety of commercial tools incorporate the manual view specification paradigm, we decided to use the less complicated visualization tool, Polestar, to control for external factors that might affect the study. The functionality of Polestar is tightly scoped and intentionaly limited to control for these potential confounds. Also, Polestar has previously been used as a control condition in previous studies~\cite{2015-voyager, 2017-voyager2}. Below, we discuss and compare the two tools.

\begin{figure}[ht]
\centering
    \includegraphics[width=\linewidth]{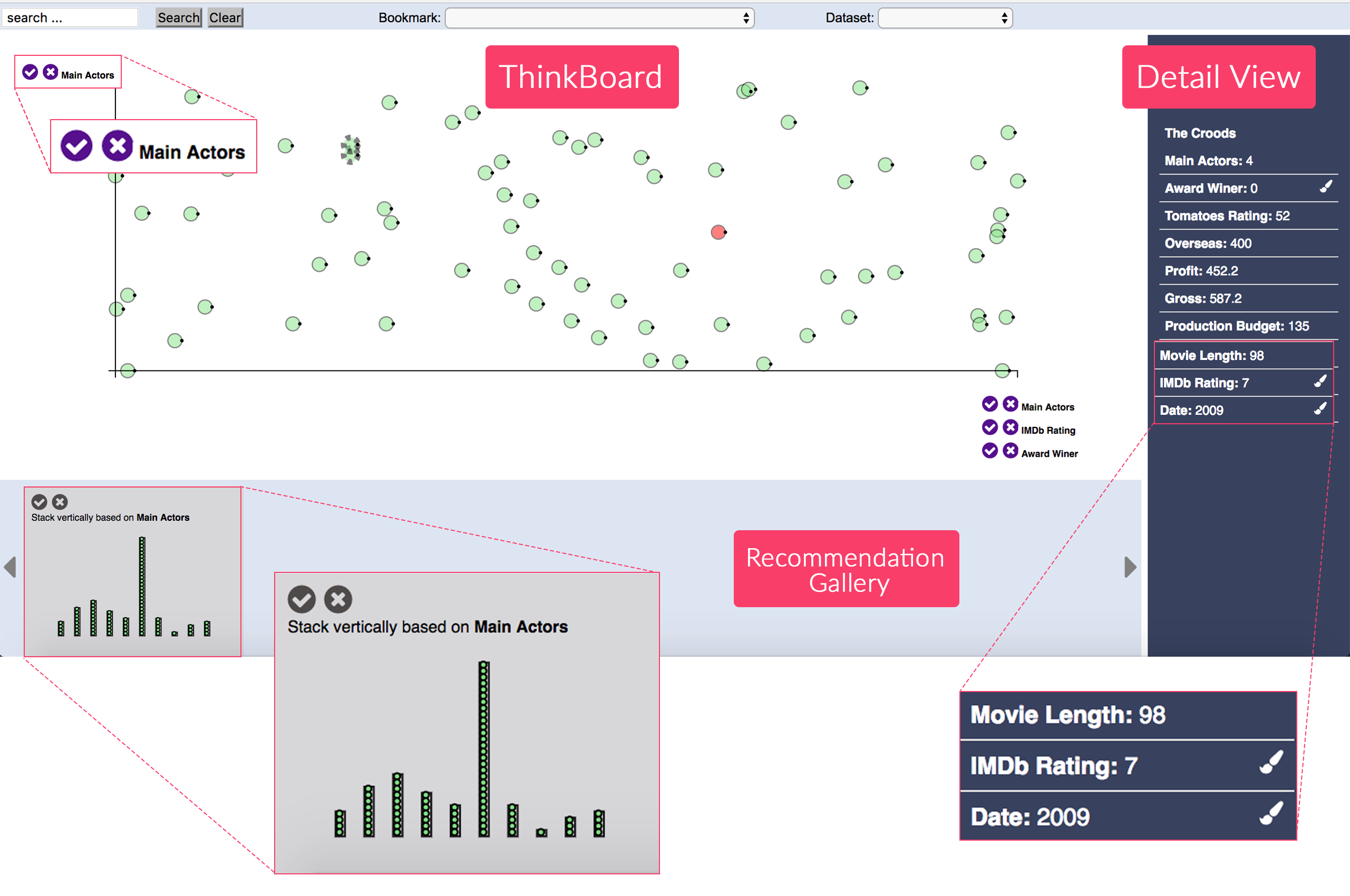} \hfill
   \caption{ A screenshot of VisExemplar, which implements Visualization by Demonstration. 
   }~\label{fig:VisExamplar} \vspace{-0.8em}
\end{figure}

\textbf{VisExemplar}\footnote{\url{https://github.com/BahadorSaket/VbD}} (shown in Figure~\ref{fig:VisExamplar}) consists of two main components: demonstrations provided by users to show their intended actions and transformations that are recommended by the system in response to the given demonstrations. To provide demonstrations, VisExemplar supports two methods. One is that it allows users to directly adjust the spatial layouts of data points (e.g., users stacking data points in the shape of bars to convey their interest in a barchart) and the other, which allows users to provide demonstrations by adjusting the graphical encodings used in a visualization (e.g., users changing the size of data points in a scatterplot). In response to the provided demonstrations, VisExemplar recommends four categories of transformations (change the current visualization technique, define mappings between graphical encodings and data attributes, assign data attributes to axes of a visualization technique, change the view specifications without changing the underlying technique). By accepting any of the recommended transformation the system will change the corresponding view.
  
The VisExemplar user interface consists of a ThinkBoard, Recommendation Gallery, and a Detail View panel. Users can construct their demonstrations through direct manipulation of the visual representation on the ThinkBoard. Some of the recommendations (e.g., recommending data attributes to be mapped to axes) might also be shown on the ThinkBoard. Other recommendations will be presented in the Recommendation Gallery. The main use of the Detail View panel is to show data attribute types and values for a selected data point. Additional details about VisExemplar and Visualization by Demonstration can be found in the previous work by Saket et al~\cite{saketVbD}.

\begin{figure}[ht]
\centering 
    \includegraphics[width=\linewidth]{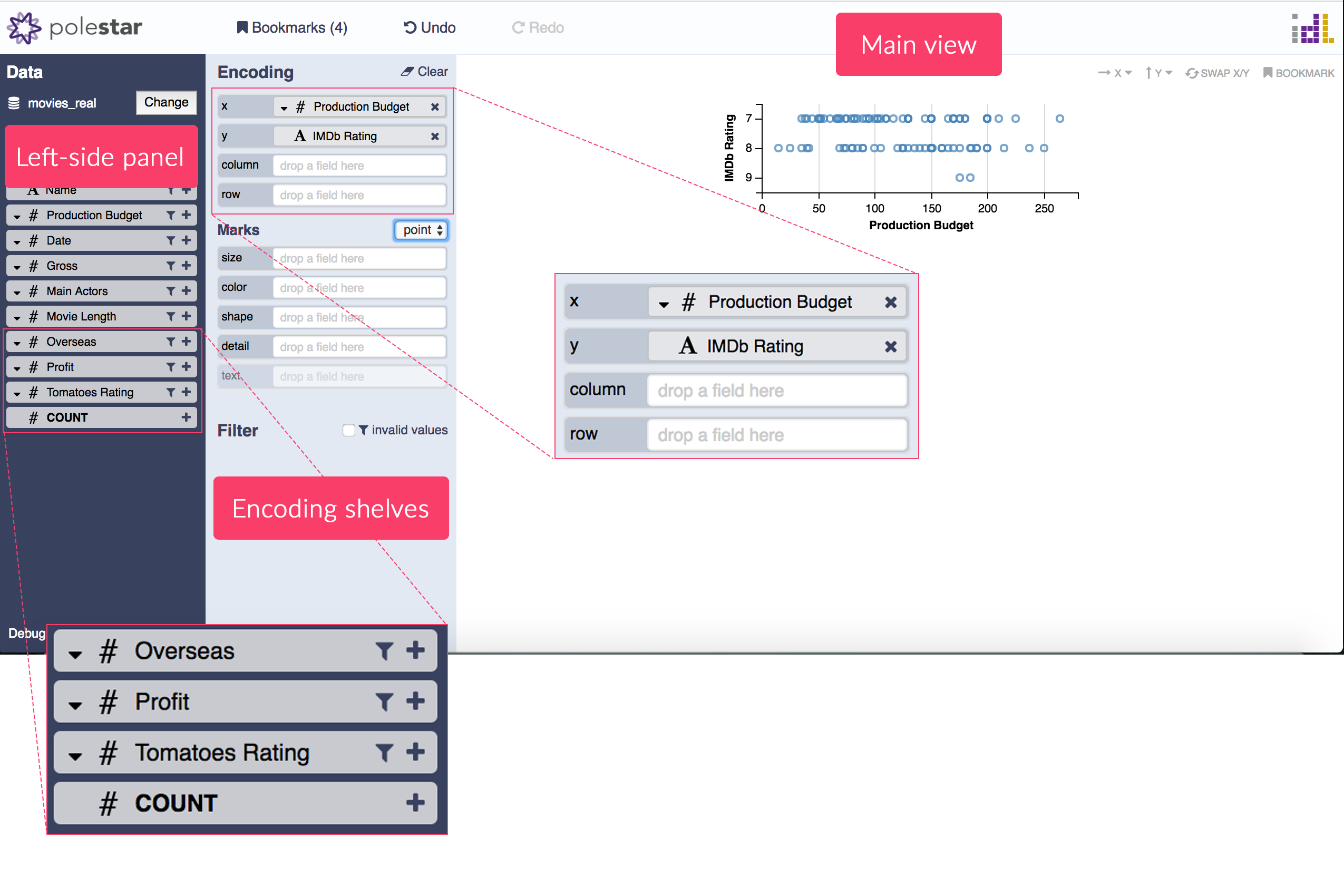}\hfill
   \caption{ A screenshot of Polestar, which implements Manual View Specification. 
   }~\label{fig:VisExamplar} 
\end{figure}

\textbf{Polestar}\footnote{\url{https://github.com/vega/polestar}} (shown in Figure~\ref{fig:VisExamplar}) is a visualization tool that implements the manual view specification paradigm. The Polestar user interface consist of a Left-side panel, Encoding Shelves, and the Main View. The Left-side panel presents the data schema, listing all data attributes in the dataset. Encoding Shelves located next to the data schema and represent different encoding channels. The main activity of designing visualizations in Polestar consists of the assignment (through drag-and-drop), of a data attribute onto a shelf to establish a visual encoding. Users can also change properties of the data (e.g., data types, data transformations) or the visual encoding variable (e.g., color or sort) via pop-up menus. The Main View of the Polestar shows a created visual representation. Upon user interaction, it updates the view accordingly. A key point here is that interactions at Polestar happen at an attribute-level.

\subsection{Similarities and Differences Between Tools}
Both tools are digital tools that share the use of marks and visual properties as main concepts in designing visualizations. They both enable users to switch between visualization techniques, map data attributes to different visual properties such as size and color, and perform operation such as sorting a bar chart and swapping the axes. However, each of these tools implements a different interaction paradigm, which results in important differences that motivate our study, as described below. \\

\noindent\textbf{Different visualization process models}

\noindent The visualization process model used in Polestar requires users to specify the data mappings prior to rendering the resulting view. In contrast, VisExamplar requires users to provide visual demonstrations of how the output of the mappings should look like. The system then interprets the provided demonstrations and recommends a set of potential mappings. A key point here is the difference in the process of mapping data attributes to visual properties in two tools.\\

\noindent\textbf{Different user and system responsibilities} 

\noindent In Polestar, users are mainly responsible for mapping data attributes to visualization properties (through drag-and-drop of data attributes to different shelves) and the system is responsible for rendering the view. In contrast, in VisExemplar, users are mostly responsible for demonstrating their intentions, and the system is responsible for interpreting users' intentions and recommending potential data mappings based on the given demonstrations. The key point here is the difference in users' and systems' responsibilities in each of these tools. \\

\noindent\textbf{Different levels of abstraction}

\noindent The main activity of constructing visualizations in Polestar involves the assignment of dragging data attributes and dropping them to encoding shelves. A main point here is that Polestar does not allow users to manipulate individual data points, but instead deal with the full range of data by attribute. Thus, interactions in Polestar happen at an attribute-level. In contrast, VisExemplar supports a lower level of   abstraction. In particular, VisExemplar enables users to provide demonstrations by manipulating the visual glyphs representing the data points. Thus, VisExemplar operates at the level of data points.



 



\section{Pilot Study}
Our study design in this paper was shaped by a pilot study. Our goal in the pilot study was to capture potential flaws in our main study design and understand if the datasets were appropriate for a 20 minutes of data exploration session during our main experiment. For the pilot study, we recruited six participants (4 male, 2 female) between 23 and 28 years old. Participants had backgrounds in computer science and social science. All of the participants had experience creating visualizations using Microsoft Excel. One of the participants also had experience in creating visualizations using Tableau. 

We randomly assigned three of the participants to work with VisExemplar and the others to work with Polestar. We first introduced each tool to the participants and trained them for 10 minutes. We then asked participants to perform eight trial tasks using the tool. The tasks were to construct and refine visualizations (e.g., sort the given bar chart in ascending order, switch between visualization techniques, assign a data attribute to the x-axis, etc.). Tasks for trial sessions were designed using dataset containing cars and various attributes describing them~\cite{henderson1981building}.

Once participants completed the tasks, we asked them to work with the visualizations tool for 20 minutes to explore a dataset about cameras~\cite{TableauData}. The participants were asked to verbalize their exploration process. We recorded 170 minutes of participants' visualization processes in the form of video screen captures. After recording the video screen captures. The analysis of the screen recordings gave us an initial understanding of the common types of specifications participants created using each paradigm and the difficulties they encountered while exploring their data using each paradigm. 

We found that Polestar worked better for cases where participants knew the exact information required for constructing or refining a visualization. In most cases, this came down to their task being defined in terms of data attributes. For example, for cases where participants knew exactly which data attribute should be mapped to which visual encoding or axis, they could do so through direct specification via the control panel. In contrast, VisExemplar worked better for cases where participants had some ideas of how the final visualization should look like, had some knowledge of the data items, but were less familiar with the data attributes or visualization terminology used in the control panel required for completing, constructing, or refining the visualization. For example, cases where the participants knew that they want a scatterplot where more expensive cameras are bigger, or even where they found specific cameras that they wanted a certain color without a formal understanding of what attributes create that mapping.

We initially decided to use the Cameras~\cite{TableauData} datasets in our main study. However, based on the pilot study, we found that participants were not familiar with many data attributes used in the Cameras dataset (or cameras in general), so they tended not to explore for as long. Participants tended not to examine attributes of the data they were not familiar with and therefore made more impetuous decisions. We instead decided to use a Movies dataset~\cite{TableauData} that provides details for 335 movies released from 2007 to 2012, and contains 12 data attributes. We selected the Movies dataset for our main experiment based on two considerations. First, the dataset contained enough data attributes to support 20 minutes data exploration. Second, the participants were unfamiliar with the content of the dataset but familiar with the meaning of the data attributes used in the dataset.

Perhaps more importantly than its results, our pilot study helped us to focus our research on a key question: How do these two interaction paradigms affect the visualization construction and visual data exploration process? To explore this question more in-depth, we conducted the following in-depth study.

\section{In-Depth Comparison of VisExemplar and Polestar}
In this phase, we conducted a think aloud exploratory observational study to understand how the participants use each interaction paradigm to construct visualizations and explore their data in a more realistic setting. We then interviewed each participant to elicit difficulties they encountered while using each paradigm. 

\subsection{Participants and Setting}
We recruited 16 participants (9 female), between 21 and 32 years old. The participants were undergraduate and graduate science and engineering students. None of them had participated in the pilot. All participants reported to be familiar with reading and creating visualizations using existing tools such as MS Excel (16), D3.js (2), and Tableau (1). None of the participants used VisExemplar or Polestar before the study. During the entire study participants used a computer with 13 inch screen. 14 of the participants never received formal instructions on visualization concepts such as visualization techniques and visual properties. 
Two of the participants took the undergraduate level information visualization course taught in our university. The study took about 1 hour to complete and participants were compensated with a \$10 Amazon gift card.


\subsection{Introduction to the Visualization Tools}
Before the main experiment, the participants were first asked to answer some demographic questions (e.g., age, sex, and prior experience in creating visualizations). We then divided the participants into two groups. One group was
introduced to VisExemplar and the other to Polestar. We walked the participants through the training session to familiarize them with the study. As our participants had no prior experience using these particular tools, we reduced their initial learning time by offering a brief introduction to the tool they would use. To prevent inconsistencies in the training session, we asked participants to watch a tutorial video of the visualization tool (we created the video prior to the study). This enabled all the participants to go through the same instruction process during the training. The video explained the fundamentals of the process of creating visual representations. The video walked the participants through different features and interactions provided by the tool. The participants were allowed to watch the video as many times as they wanted. After watching the video, we asked participants to work with the tool for 15 minutes using the Cars dataset. We encouraged the participants to ask as many questions as they want during this stage. 

To ensure participants familiarity with both tools, we asked participants to perform 24 tasks (4 categories of tasks $\times$ 6 trials).  All tasks were printed on a sheet of paper. Each time the interviewer selected a task randomly and asked the participants to perform the task. The participants were not allowed to move to the next training question unless they answered the question correctly. Once comfortable with using the visualization tool, users were instructed to take a short break and move to the main study.

To select tasks for training the participants, we first interacted with both VisExemplar and Polestar for a week exploring different ways in which they support visualization construction and reconfiguration. This resulted in a list of 45 visualization construction tasks. We also reviewed taxonomies of tasks commonly used for interactive visualization construction and data exploration (e.g.,~\cite{Shneiderman:taxonomy, yi2007toward, Ren:taxonomy, Dix:taxonomy}). Considering our experiences with these tools and our knowledge from these taxonomies, we then assigned these tasks into four categories according to the type of changes they make to a visualization. \\

\begin{itemize}[leftmargin=4mm]
  \item \textbf{Mapping data attributes to the axes:} It requires users to assign data attributes to either one or both axes of a visualization. \\
  
  \item \textbf{Mapping data attributes to visual encodings:} It requires users to map a data attribute to a visual encoding. \\
  
  \item \textbf{Switching between visualization techniques:} It requires users to change from one visualization technique to a different visualization technique. \\
  
  \item \textbf{Reconfiguring a visualization:} It requires users to change the view specification of a visualization without changing the underlying technique and mappings.\\ 
\end{itemize}

At the end of this phase, all participants successfully completed these tasks, and were comfortable with the functionality of their tool.

\subsection{Main Study Procedure} 
After introducing the tools to the participants and training them (section 6.2), the participants were asked to explore the Movies dataset~\cite{TableauData} and look for interesting findings about the data. In particular, the participants were told to imagine their employer asked them to analyze the dataset using the visualization tool for 20 minutes and report their findings about the data. The participants were instructed to verbalize analytical questions they have about the data, the tasks they perform to answer those questions, and their answers to those questions in a think-aloud manner. In addition, we instructed them to come up with data-driven findings rather than making preconceived assumptions about the data. The participants were not allowed to ask any question during this phase. We tried to avoid interrupting the participants as much as possible during their data exploration process. However, we sometimes needed to remind the participants that this is a think-alound study and they need to verbalize their thoughts.

This phase of our study concluded with a follow-up interview, in which we asked participants about what they liked and disliked about the interaction paradigm. This was to allow the participants to convey their feedback and ideas and in order
to solicit potentially unexpected insights. We asked participant to explain what they liked or disliked about the interaction paradigm implemented in the visualization tool they used. We instructed the participants to give feedback on the interaction paradigm incorporated in the tool rather than the user interface. 

\subsection{Data Collection and Analysis} 
To analyze differences between the VisExemplar and Polestar conditions, we gathered several types of data. At the beginning of the training session, we used questionnaires to collect participant demographic and background information. During the main study, we took written notes of participants' interaction processes with the tools. We also recorded the participants' audio and the screen of the computer they worked with.

To analyze the video and interview material, we followed guidelines provided by Creswell~\cite[p.~236]{creswell2002educational} for analyzing qualitative data. We first fully transcribed data from the interviews. The coder (first author) then read the transcribed materials to obtain a general sense of the data and thinking about organization of the data. After reading the data, the first author identified the meaningful text segments and assigned a code word or phrase that accurately describes the meaning of the text segment. The coding process was an iterative process with three passes by a single coder in which the coder developed and refined the codes. During the coding phase, we mainly focused on processes of the participants in terms of \textbf{usage} (what types of visualization specifications were usually created using each interaction paradigm? What are the patterns in the use of specific functionality?) and \textbf{barriers} (when and how difficulties happened while working with each paradigm?) For example, our codes included phrases such as ``changing color'', ``changing size'' and ``stacking data points''.  
Finally, we identified and aggregated similar codes that frequently occurred to form themes. Similar to codes we assigned labels to themes. For example, we aggregated the ``changing size'' and ``changing color'' codes to create ``mapping a data attribute to a visual property'' theme. Finally, we identified frequently occurring codes and themes to form higher-level descriptions and to discuss our findings. 



\section{Findings}
In this section, we categorize and discuss the findings of our study. 

\subsection{No-need-to-think vs. Need-to-think}
Our data analysis revealed that the fundamentally different interaction paradigms represented by these two tools have an effect on how participants carried out their visualization construction process and explored alternative designs. Participants following Polestar's design lead in the absence of an initial idea or reflection on adequate data mappings. To create different visualizations in Polestar, the participants constantly assigned data attributes to different visual properties through GUI operations visually presented via the control panel. 
While participants experienced fast data exploration using Polestar, they usually did not reflect much on the meaning and potential impact of their exploration. For example, one of the participants stated: \textit{``I did not need to think of how I want to create visualizations. When I started, I did not have any design in my head. So, I kept specifying attributes and creating different designs until I found the one [visualization] that looked interesting.''} Another participant mentioned: \textit{``You know, for example here, initially, I did not plan to create this scatterplot with profit on the x-Axis, but after trying multiple data attributes, I felt profit might be an interesting one to look at.''} In such cases participants tried mapping a variety of data attributes to different visual properties until they created a visualization that they liked. 

Previous work~\cite{Mendez:2017} also confirms this notion of ``no-need-to-think'' when working with Tableau (another tool implementing manual view specification). This notion enables users to quickly explore a variety of visualization alternatives and explore their data. In fact, results of our interviews indicate that many of the participants were satisfied that they could map data attributes to different visual properties quickly in Polestar. They found it particularly effective when they had no plan about the type of visualization they wanted to create and just wanted to try different visualization alternatives.

In contrast, VisExemplar advocates for the notion of ``need-to-think''. Users need to think about the visual output or how data mappings should look before starting the demonstration process. With VisExemplar, five participants mentioned that they had to think about how they want their visual outputs to look and only then started providing visual demonstrations of incremental changes to the visual representation. One of the participants stated: \textit{``Here [in VisExemplar] I need to think and imagine the output first. I then need to come up with strategies to show [demonstrate] parts of what I want to the system.''} 
Another participant highlighted the need for imagining the visual output before starting the data mapping by saying: \textit{``[...] You come up with the plot in your mind, then demonstrate what you want to the system, and let the system give you a set of potential options.} What we can draw attention to here is that unlike Polestar, VisExemplar advocates for the notion of ``need-to-think" before specifying visual properties. 

The notion of ``need-to-think'' in VisExemplar obviously requires users to put more effort into both thinking about their visualization designs and providing demonstrations to the system, thus making the processes of visualization construction and visual data exploration slower. 
However, this notion might lead to a more thoughtful process since the participants often mentioned the need to plan and/or think prior to engaging in the process of visualization construction. In particular, the notion of ``need-to-think'' might enable users to think more carefully about marks, visual variables, and their relevance in the design and construction of visualizations.

\subsection{Pre-defined Controls vs. Free-Form Interaction }
Polestar enables rapid and formal/exact specification of the visualization properties by incorporating GUI elements with a pre-defined set of controls. Thus, the way that people construct visualizations and explore their data is based on the actions provided by these controls in the interface. This is consistent with past literature describing how people form goals from available affordances~\cite{suchman1987plans, shneiderman1994dynamic}. While this design increases the speed of the data exploration process and the number of visualizations created by the users, it limits flexibility in the visualization construction process. In contrast, VisExemplar enables users to construct visualizations through more free-form interactions. It enables users to interact at the data point level to demonstrate their intentions, manipulating the data points in a variety of ways that lead to a potentially large set of possible visualizations. This degree of flexibility in interaction has an impact on visualization construction, as suggested by our findings below.

In Polestar, participants based their construction process on the affordances of the controls provided in the interface. For example, one of the participants emphasized this point by saying: \textit{``I think the process [manual view specification] emphasizes on trying different attributes and checking different plots. [...], I mainly played with options given to me on this [control] panel to change different encodings.''} Another participant stated: \textit{``It [Polestar] enables you to use a set of templates and create a bunch of visualizations quickly.''} This might suggest that using pre-defined controls is well suited for rapid visual data exploration where the emphasis is more on constructing a large number of visualizations to look at different aspects of data.

\begin{figure}
\centering
 \centering
  \includegraphics[width=\columnwidth]{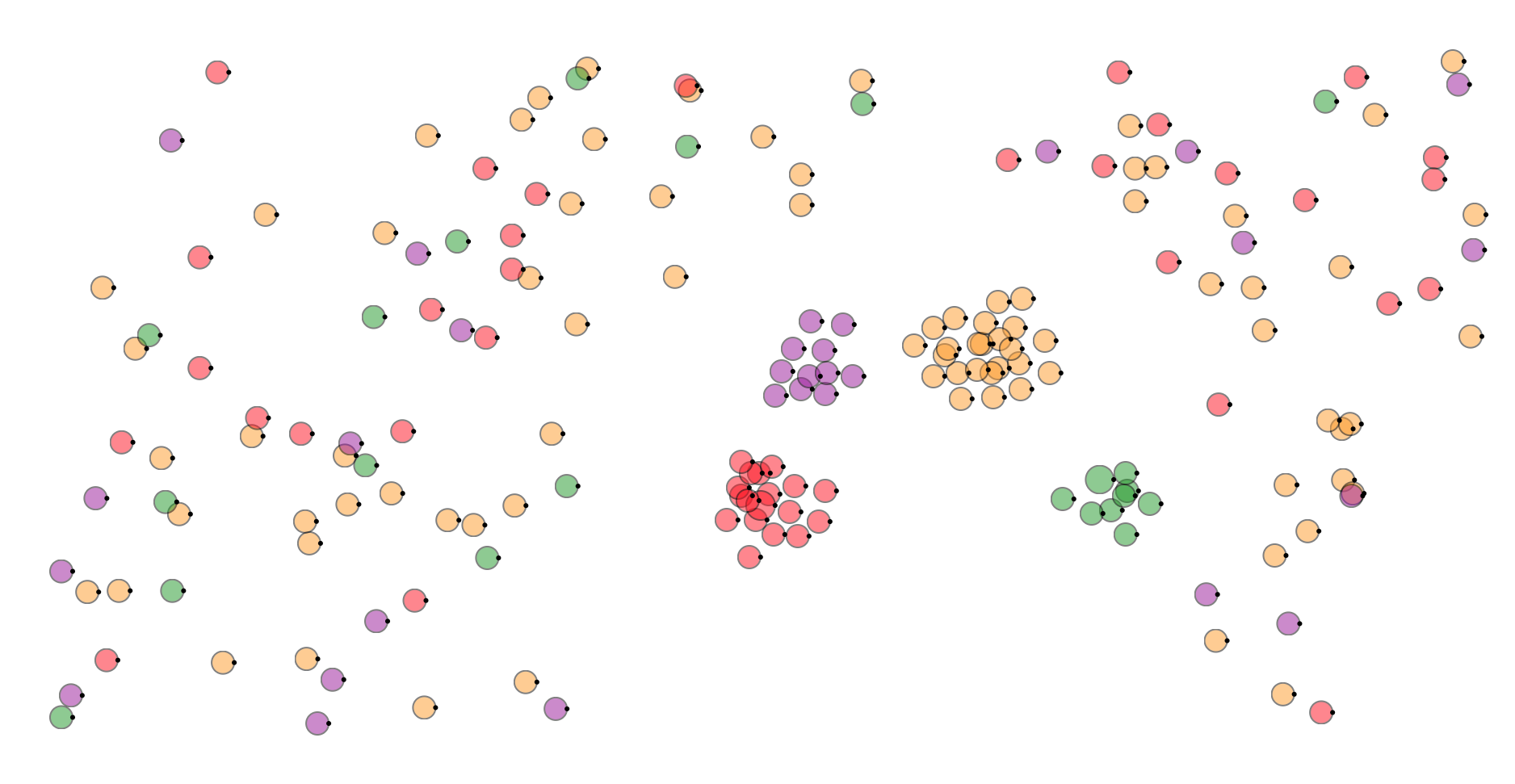}
\caption{A user demonstrates his interest in creating a visual design where movies with the same genre fall in the same group/cluster.}~\label{fig:grouped}  \vspace{1.2em}
\end{figure}

\begin{figure}
\centering
 \centering
  \includegraphics[width=\columnwidth]{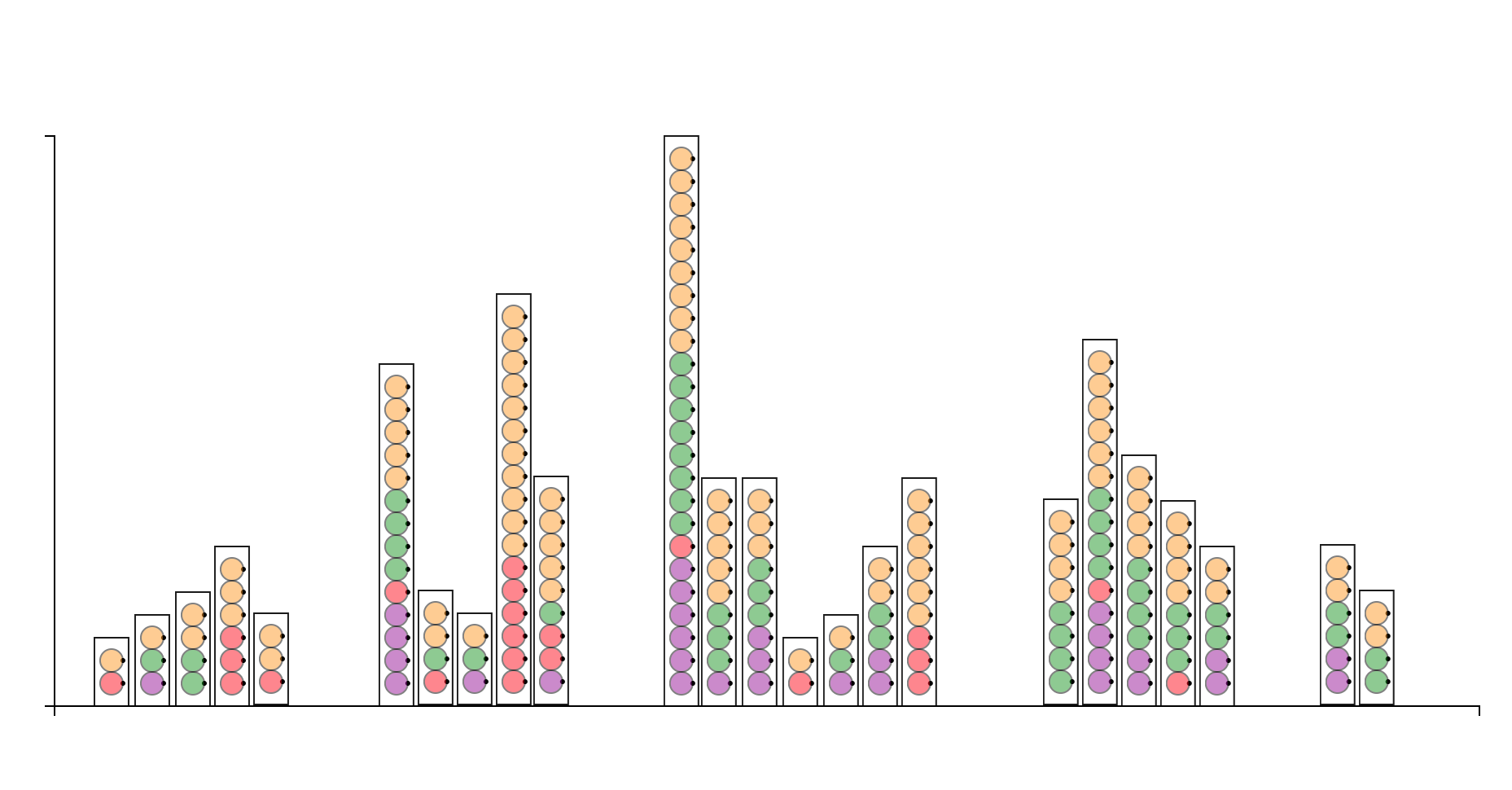}
\caption{A user demonstrates his interest in binning the x-axis of the visualization by grouping the bars spatially. }~\label{fig:binning}  \vspace{1.2em}
\end{figure}

The free-from interaction in VisExemplar sometimes led users to be more creative and demonstrate their interest in designing novel visualizations (some of which are not even supported by VisExemplar). 
For example, one of the participants said: \textit{`The system allows me to interact with the elements on the screen to create visualizations. I can assemble data points.''} We also observed the intentions in creating novel visualizations in some of the demonstrations provided by the participants. For example, one of the participants expressed his interest in creating a view where data points are grouped (See Figure~\ref{fig:grouped}). Another participant demonstrated his interest in binning the x-axis (See Figure~\ref{fig:binning}). The same participant states: \textit{``For me the length of a movie is an important factor. I want to divide the movies' length into five groups. This way I can just focus my exploration on movies in a specific length range.''} These findings suggest that visualization by demonstration helps people experiment with many alternative visualization designs. In contrast to pre-defined control panels, the affordances of demonstration-based interfaces allow people to translate their visual goals into demonstrations, potentially broadening the set of mappings and visualization techniques considered and used.\\

\subsection{Perceived Control Over Visualization Construction}
The level of control users perceived over the visualization construction process is different in the two tools studied. Polestar uses a conversation metaphor~\cite{hutchins1985direct}, in which it introduces GUI elements such as menus and shelves that act as mediators between users and the visual representation. In this case, the interface is an intermediary between the users and the visual representation. Using Polestar, one of the participants felt a lack of control over the construction process by saying: \textit{``I feel like my interaction with the system is not very explicit.''} Another participant stated: \textit{``[...] the process of making graphs is automated here [in Polestar]. I drag an attribute and drop it here [shelves], and the system creates the graph.''}

In contrast, VisExemplar mainly relies on a model-world metaphor~\cite{hutchins1985direct}, in which the interface is itself a visual representation and the user can act on the visual representation rather than external interface elements. In VisExemplar, users can directly manipulate the graphical encodings used in the visual representation. The level of interaction directness~\cite{beaudouin2000instrumental} with the visual representation contributes towards increasing the perceived control of the participants over the visualization construction process. While using VisExemplar, three of the participants commented on their level of control over the visualization process that resulted from their freedom in interacting with graphical encodings. For example, one participant mentioned: \textit{``Being able to demonstrate to the system what I want gives me control over what I am doing. What I mean is that I can do what I want with these circles. I can move them, color them, resize them.''} Another participant stated: \textit{``I like that I can color it here [coloring the glyph], I feel like I have control over the circles [data points]''.} 
Therefore, the choice of interaction paradigm can change the way that people perceive their role (or control) in the visualization construction process.

\subsection{Influence on Visualization Specification}
Below, we divided types of specifications the participants created during the entire visualization construction process into four categories: mapping data attributes to axes, mapping data attributes to encodings, switching between visualization techniques, and reconfiguring a visualization. We discuss advantages and disadvantages of each tool while creating these specifications. \\

\noindent \textbf{Mapping Data Attributes to Axes} 

\noindent Participants tended to map more data attributes to axes using Polestar (see Table~\ref{TAB:operationNumber}). Four of the eight participants who worked with Polestar stated that the fast speed of the tool in mapping data attributes to axes have contributed to this advantage. For example, one of the participants mentioned \textit{``It is easy to drag and drop data attributes in this tool [Polestar]. I can quickly change axes.''} On the other hand, in VisExemplar, five of the participants expressed difficulties in mapping data attributes to axes. To map a data attribute to an axis using VisExemplar, the participants had to position a few data points relative to their data attribute values. The system then recommended potential data attributes to be assigned to the axes. For example, one participant expressed how a large amount of effort was required for him to map a data attribute to an axis: \textit{``I need to drag a point, keep track of its value, and compare its position with other points.''} Later, during the follow-up interview, the same participant mentioned: \textit{``You know it is hard to drag the points and track their values, [...] maybe you could somehow highlight the values [data attribute values] while moving the points to decrease users' cognitive load.'' } might suggest that for tasks where the attributes are known, creating mappings for them is more easily done with an interaction paradigm such as the one implemented by Polestar.

Another challenge that a few of the participants encountered while using VisExemplar was the accuracy of the data attributes suggested to be mapped to the axes. After a position-changing interaction, the system searches for data attributes to recommend for mapping to the axes based on the positions of the moved data points. The recommendation engine prioritizes potential suggestions and shows those above a certain threshold. However, there might be cases where a user's expected data attribute is not among those recognized to be the most related ones by the system. In such cases, users have to provide more demonstrations to help the system to interpret their intentions better. However, providing more demonstrations can be frustrating in some cases. One of the participants mentioned her concern by saying: \textit{``The recommendations on the axes don't always make sense to me. When I have an idea in mind like let me see how these [data attributes] compare, then when I don't see it in the options, I am kinda thrown off because at that point I am kinda doubting whether the way that I am thinking about it is wrong or whether I am doing something wrong with the system.''} \\

\noindent \textbf{Mapping Data Attributes to Visual Encodings} 

\noindent The participants mapped more data attributes to visual encodings using VisExemplar (see Table~\ref{TAB:operationNumber}). To map a data attribute to visual encoding using VisExemplar, users could directly manipulate characteristics (e.g., size) of a corresponding encoding in the visual representation. For example, users could color one or more data points red to convey their interest in mapping this specific color to a data attribute. The system then recommends a set of data attributes that can be mapped to color. During data analysis, we noted three interesting findings in participants' exploration process.

First, in VisExemplar, users could color or resize data point(s) and the system would then recommend appropriate data attributes that they could assign to color or size. Participants found the process of recommending a subset of data attributes for color and size very interesting and helpful. For instance, one participant mentioned that: \textit{``[...] coloring points was fast though. I can color one point and the system suggests a small set of attributes.''} Another participant expressed this by saying: \textit{``it [VisExemplar] makes coloring easy for me because suggesting attributes helps me to focus on a subset of attributes.''} On the other hand, one of the participants who used Polestar stated: \textit{``every time I need to skim through attributes on this panel [the panel showing data attributes], pick one, and drag it. It becomes hard to skim through all attributes if we have many of them [data attributes].''} 

Second, we saw an interesting pattern emerge when the participants did not intend to map any specific data attribute to an encoding but wanted to explore different mapping options by hovering on the recommended data attributes. For example regarding VisExemplar, one of the participants mentioned: \textit{``... let's color one and look at recommendations [recommended data attributes].''} This is interesting because if the participants wanted to do this using Polestar, they would have to drag and drop a data attribute to an encoding every time to see different options. However, this was different in VisExemplar. The participants could color or resize a data point to get data attributes recommended to them by the system. They then would hover on those recommended attributes to preview the results and explain their findings. Six of the participants really appreciated the capability of the system in recommending data attributes and enabling users to quickly preview the mappings. For example, one participant who had experience using other visualization tools mentioned: \textit{``[...] in tools that I used for the InfoVis class, we had to assign an attribute to color for each attribute separately, but here [using VisExemplar] it is easier because I can color one circle [data point] and the system gives me a subset of attributes, and I can hover on them to see [preview] the results.''} 

Third, the participants felt more control over the tool when they were mapping data attributes to visual encodings using VisExemplar. This is potentially because the tool enables them to directly manipulate characteristics (e.g., size or color) of a glyph. This way the interaction between the users and the system is ``direct''~\cite{beaudouin2000instrumental} and there is no intermediary between the users and the visual representation (e.g., users do not need to use a control panel). One participant expressed his feeling of having control by saying: \textit{``I like that I can color it here [coloring the glyph], I feel like I have control over the circles [data points]''.} \\

\begin{table}
\footnotesize
  \caption{This shows the total and average number of times that participants created each type of visualization specification using VisExemplar and Polestar.} \label{TAB:operationNumber} 
   \begin{tabular}{p{0.2\textwidth}p{0.03\textwidth}p{0.07\textwidth}p{0.08\textwidth}}
    \toprule
      {\sc Specification Type} & & {\sc VisExemplar} & {\sc Polestar}\\
    \midrule
      Mapping data attributes to axes  & Total & 53 & 108 \\
      & Avg & 7.5 & 15.4 \\
      
   \midrule   
       Mapping data attributes to visual encodings  & Total& 55 & 26 \\
       & Avg & 3.9 & 1.9 \\
    \midrule    
        Switching between visualization techniques & Total& 12 & 25 \\
       & Avg & 1.4 & 3.5 \\
    \midrule   
          Reconfiguring a visualization & Total & 7 & 9 \\
      & Avg & 0.8 & 1 \\
    
    \bottomrule
 \end{tabular} 
\end{table}

\noindent \textbf{Switching Between Visualization Techniques} 

\noindent The participants switched between visualization techniques more often while using Polestar (see Table~\ref{TAB:operationNumber}). The ability to quickly change from one type of technique to another could contribute to this advantage. In particular, seven of the participants found it quite difficult to switch from a scatterplot to a barchart using VisExemplar. 
To switch from a scatterplot to a barchart using VisExemplar, the participants had to stack two or more data points vertically. The system then recommended a set of barcharts based on similarity of the data points. The participants found stacking data points difficult. One participant expressed the difficulty of switching from a scatterplot to a barchart by saying: \textit{``it was a bit awkward for me to stack the points to create a barchart''} \\

\noindent \textbf{Reconfiguring a Visualization} 

\noindent We did not find a large difference in the number of times that the participants reconfigured the visualizations using each tool. In fact, the results of our first phase also indicate that both tools were quite fast in reconfiguring visualizations. However, we noted four of the participants found sorting the bar chart using VisExemplar intuitive and fun. For instance, one of the participants mentioned: \textit{``[...] the sorting was intuitive.''} Another participant said: \textit{``interactions like sorting are fun and natural. Have you ever thought to test your tool on high school students? I think they will like it a lot because they can move things around and play with it while they are learning.''} VisExemplar enables users to demonstrate their intrests in sorting the bar chart by dragging the shortest/tallest bar to the extreme left/right. The system then recommends sorting the bar chart.

\subsection{Different Strategies for Phrasing Goals}
We captured two strategies (\textit{\textbf{specific}} and \textit{\textbf{abstract}}) that the participants used to phrase their tasks while constructing or refining a visualization. In the specific strategy, the participants knew the exact information required for constructing or refining a visualization. For example, they knew which data attribute should be mapped to which visual property. While using Polestar, participants often used the specific strategy to explain their goals/tasks. For example, participants mentioned: \textit{``I want profit on the x-Axis.''}, \textit{``I am sorting the bar chart in ascending order.''}, \textit{``I am mapping Genre to the color encoding.''} 

The second strategy of phrasing goals/tasks is more abstract. While using VisExemplar, in some cases participants were unaware of some of the information required for completing, constructing, or refining a visualization. So, they phrased their goal in a more abstract way. For example, the participants sometimes tried to express that they wanted to map an attribute to size by saying, \textit{``I want to make long movies bigger.''}. Another participant phrased her interest in assigning rating to the x-axis of a scatterplot by saying: \textit{``I am changing the representation so that circles are horizontally positioned by rating''}. We call this an abstract strategy because the participants had partial information about the details of the intended specifications (often missing either the exact attribute needed, or the name of the command to create the mapping). Instead, they demonstrated a part of the mapping or result and let the system figure out what is a good data attribute.  

The key difference between these two strategies is how well the users can articulate their goals based on the data attributes and visualization characteristics. When goals are formed in the way of data attributes and visual mappings, the specific strategy was used. Alternatively, when goals are formed on data items and semantic relationships between data items, abstract strategies were used.

\section{Discussion}
In this section, we discuss the trade-offs between the interaction paradigms implemented in two tools.  

\subsection{Manual View Specification}
All the participants easily learned and followed the manual view specification's design. Manual view specification is a fast and accurate paradigm since it enables rapid and formal/exact specification of the visual properties by incorporating a set of consistent user interface elements. Tools implementing the manual view specification support the notion of ``no-need-to-think'', which is particularly effective when users have no visual design in their minds and want to explore their data by creating as many visualization alternatives as possible in a limited time.

One minor challenge in interacting with tools such as Polestar is that they do not enable users to preview views as a result of data mappings. For example, Polestar requires users to first specify the mappings, the system then renders the view. A few participants raised this challenge in Polestar. They mentioned wanting a feature that enabled them to preview the resulting views before committing. We envision several ways that tools implementing manual view specification can incorporate previewing resulting views. For example, in Polestar, imagine participants could first select/lock a visual property (e.g., size). They could then hover on different attributes to see the outcome of mapping different data attributes to that specific visual property.

\subsection{Visualization by Demonstration}
We found that visualization by demonstration emphasizes on the notion of ``need-to-think'' that requires users to think about their designs before starting visualization construction. Flexible and free-form nature of interactions in visualization by demonstration enabled users to be more creative and think of visualizations that are novel and in some cases are not even supported by the system. That being said, participants raised multiple challenges in visualization by demonstrations that we discuss below.\\

\noindent\textbf{Advancing Interactions}

\noindent Providing demonstrations is one of the fundamental steps in the visualization by demonstration paradigm. Participants encountered two types of challenges while providing visual demonstrations. First, providing certain demonstrations requires time and accuracy. For example, to switch from a scatterplot to a barchart, participants needed to stack a few data points vertically so that data points overlapped. This takes participants time to demonstrate because they need to drag at least two data points and stack them. Second, providing certain demonstrations is cognitively complex. For instance, mapping data attribute to axes requires participants to simultaneously keep track of a data point values and their position relative to other points. 

Going forward, we envision multiple ways to avoid these interaction challenges in tools implementing visualization by demonstration. One way is to incorporate more advanced interactions such as lasso selection to improve users' speed in providing multi-point demonstrations. For example, imagine switching from a scatterplot to a bar chart. Users could do a lasso selection and then a mouse gesture to stack them. Another potential solution might be to incorporate feedforward~\cite{Vermeulen} and suggested interactivity~\cite{boy:hal-01188973} to improve the efficiency of visualization by demonstration. For tasks such as ``mapping data attributes to the axes'' this could help by showing what sequences of operations a user should execute to provide a demonstration. Finally, systems using visualization by demonstration could incorporate methods for making the affordances more salient to the users. For example, they could provide an animation upon hovering on each encoding to show how to interact with it. \\

\noindent\textbf{Improving Recommendations}

\noindent While using VisExemplar, there were cases where the participants were unclear why the system suggested specific recommendations. In such cases, the participants found it difficult to map the recommended options to their interaction with the visualization. For example, one of the participants stated: \textit{``If the system could explain to me why it is suggesting me these bar charts, I would then adjust my actions to get more related suggestions.''} All recommendation systems suggest potential options to users based on a specific set of criteria. There might be cases that the systems do not recommend options expected by the users, and it might not be apparent to users why those recommendations are presented to them. Going forward, we suggest the systems implementing visualization by demonstration to design methods to explain the reasoning behind recommendations. Thus, one of the open challenges in this line of research is how to explain the recommendations.

We also noticed that the participants sometimes found incoming recommendations interrupting. For example, one of the participants mentioned that \textit{``is there a way to tell the system to do not update the recommendations after each interaction?''} In the current version of VisExemplar, the recommendations will be updated in the interface whenever the recommendation table in the recommendation engine gets updated. We suggest the systems which plan to make use of the visualization by demonstration paradigm consider investigating methods for minimizing the interruption caused by incoming recommendations. We can envision two strategies to overcome the timing problem. First, systems present recommendations upon pressing a specific button on the interface. Second, systems could observe the cadence of user interaction with the system and make recommendations at a less active time.


\section{Limitations and Future Work}
This study is a first attempt to investigate people's visualization construction using manual view specification and visualization by demonstration. However, a single study cannot answer all open questions about this process and is necessarily limited by several factors. In this section, we discuss the limitations in our comparison of manual view specification and visualization by demonstration.

\subsection{Implementation and Design of Visualization Tools}
Our findings should be interpreted in the context of the specified visualization tools. To compare two interaction paradigms, we had to select two digital tools that implement one of the interaction paradigms. Thus, we chose VisExemplar and Polestar because each implements one of the interaction paradigms and their relative simplicity with regards to the remaining system components. In this paper, we mainly discussed the results that highlight advantages and disadvantages of the underlying interaction paradigms rather than a specific implementation of each tool. However, we want to emphasize that tool design and implementation might influence the results of the study. Thus, we encourage future work to consider the effect of tool design and implementation when exploring our findings.

\subsection{Dataset, Tasks, and Visualization Techniques}
This study is a non-exhaustive examination of manual view specification and visualization by demonstrations. Tools studied in this paper are only used to generate a subset of visualization techniques (bar chart and scatterplot), based on only one type of dataset (tabular datasets). Our findings should be interpreted in the context of the specified visualization techniques, interactions, and tasks. Testing our research questions using more advanced visualization tools, more sophisticated tasks, dataset types, and visualization techniques (e.g., sunburst) might reveal nuances of each interaction paradigm that were not tested in this study. Moreover, this study can be extended with other studies such as comparing these two paradigms on users with different level of visualization expertise (novel vs. expert) and testing other interaction modalities (e.g., touch, speech, etc.).

\section{Conclusions}
We present a study comparing people's visualization process using two visualization tools: one promoting manual view specification (Polestar) and another promoting visualization by demonstration (VisExemplar). Our findings indicate differences in how these interaction paradigms shape people's visualization construction and data exploration processes, people's decision and strategies for visualization design, and people's experience and their feeling of control and engagement during visualization process. We also discuss some of the trade-offs and open challenges in incorporating these interaction paradigms.

\bibliographystyle{abbrv}

\bibliography{template}

\begin{IEEEbiographynophoto}{Bahador Saket}
is currently a Ph.D. student at Georgia Institute of Technology. His research areas of interest include human-computer interaction and information visualization.
\end{IEEEbiographynophoto}

\begin{IEEEbiographynophoto}{Alex Endert} is an Assistant Professor in the School of Interactive Computing at Georgia Institute of Technology. He directs the Visual Analytics Lab, where him and his students explore novel user interaction techniques for visual analytics. 
\end{IEEEbiographynophoto}

\end{document}